\documentclass[12pt]{article}
\usepackage{amsfonts}
\usepackage{color}
\newcommand{\Section}[1]%
{\section{#1}\setcounter{equation}{0}%
\setcounter{theorem}{0}}
\newtheorem{theorem}{Theorem}
\newtheorem{lemma}[theorem]{Lemma}
\def\re{\mathbb{R}}

\def\ze{\mathbb{Z}}

{\par\noindent{\em #1:\ }}%
{~\rule{2mm}{2mm}\par\bigskip}
\begin{document}
%%%%%%%%%%%%%%%%%%%%%%%%%%%%%%%%%%%%%%%%%%%
%%%%%%%%%%%%%%%%%%%%%%%%%%%%%%%%%%%%%%%%%%%
%%%%%%%%%%%%%%%%%%%%%%%%%%%%%%%%%%%%%%%%%%%
\newpage\thispagestyle{empty}
{\topskip 2cm
\begin{center}
{\Large\bf Decay of Superconducting Correlations for  
Gauged Electrons in Dimensions $D\le 4$\\} 
\bigskip\bigskip
{\Large Yasuhiro Tada\footnote{\small \it Institute for Solid State Physics, The University of Tokyo, 
Kashiwa 277-8581, JAPAN, 
{\small\tt e-mail: tada@issp.u-tokyo.ac.jp}}$^{,}$
\footnote{\small \it 
Max Planck Institute for the Physics of Complex Systems,
N{\"o}thnitzer Str. 38, 01187 Dresden, GERMANY}
and Tohru Koma\footnote{\small \it Department of Physics, Gakushuin University, Mejiro, Toshima-ku, Tokyo 171-8588, JAPAN,
{\small\tt e-mail: tohru.koma@gakushuin.ac.jp}}
\\}
\end{center}
\vfil
\noindent
{\bf Abstract:} We study lattice superconductors coupled to gauge fields,   
such as an attractive Hubbard model in electromagnetic fields, with a standard gauge fixing. 
We prove upper bounds for a two-point Cooper pair correlation at finite temperatures in spatial dimensions $D\le 4$. 
The upper bounds decay exponentially in three dimensions, and by power law in four dimensions.
These imply absence of the superconducting long-range order for the Cooper pair amplitude 
as a consequence of fluctuations of the gauge fields. 
Since our results hold for the gauge fixing Hamiltonian, they cannot be obtained as a corollary of Elitzur's theorem. 
\par
%%%%%%%%%%%%%%%%%%%%%%%%%%%%%%%%%%%%%%%%%%%%%%%%%%%%%%%%%%%%%%%%%%%%%%%%%%%
\noindent
\bigskip
\hrule
\bigskip
%%%%%%%%%%%%%%%%%%%%%%%%%%%%%%%%%%%%%%%%%%%%%%%%%%%%%%
\vfil}
%\newpage
%%%%%%%%%%%%%%%%%%%%%%%%%%%%%%%%%%%%%%%%%%%

%%%%%%%%%%%%%%%%%%%%%%%%%%%%%%%%%%%%%%%%%%%%%%%%%%%%%%
\Section{Introduction}
For understanding superconductivity, taking into account electromagnetic fields is indispensable. 
Actually, the Meissner effect is the expulsion of external magnetic fields from the bulk region of a superconductor.
For describing the electromagnetic fields, Maxwell equations of classical electromagnetism 
have been often used in theoretical approaches, which are based on 
Ginzburg-Landau (GL) or Bardeen-Cooper-Schrieffer (BCS) theories~\cite{Schrieffer,Tinkham}.  
Namely, quantum and/or thermal fluctuations of the electromagnetic fields have been often ignored because 
detecting evidence of the effect of the fluctuations is fairly difficult in experiments of 
superconductivity. 

On the other hand, it is widely believed that the Meissner effect is described as a Higgs phenomenon 
where the gauge fields dynamically gain a mass even without U(1) symmetry breaking~\cite{FS,FMS,BN1,BN2}.
Such a phase is proposed to be understood as an intrinsic topological order 
which is characterized by a gauge theory~\cite{FS,HOS}. 
If the standard Coulomb gauge condition which yields transverse photons only
is imposed without taking into account the fluctuations of the gauge fields, 
then it seems very hard to explain the emergence of
the massive photons which necessarily have a longitudinal component 
by the standard theory of massive vector fields. 
Thus, the issue of the fluctuations of the gauge fields is
significantly important for superconductivity, and a superconductor in fluctuating electromagnetic fields should be modeled 
by charged fermions coupled to a dynamical U(1) gauge field. 
According to the well-known gauge principle, a physical system must be invariant under local gauge transformations. 
For such a model, Maxwell equations of classical electromagnetism can be derived 
as a result of a saddle-point approximation for the gauge field.  
However, it is considerably hard to calculate generic physical quantities for a fully gauge invariant system 
by going beyond such approximations. In fact, there arise some problems,  
e.g., gauge redundancy for perturbative approaches and complex U(1) phases of fermion hopping 
for Monte Carlo calculations. 

Besides, under the assumption of the gauge principle, Elitzur's theorem~\cite{Elitzur,DdFG} states that 
if a local observable has a nontrivial representation under local gauge transformations, 
then the expectation value of the observable is necessarily vanishing. 
This implies that the Cooper pair amplitudes in superconductors are necessarily vanishing \cite{TadaKoma}, 
when taking into account the fluctuations of the electromagnetic fields. 
In order to avoid this difficulty, gauge fixing conditions have been widely used 
for calculating physical quantities in both compact and non-compact gauge theories~\cite{DZ,LNW}.
In particular, Kennedy and King \cite{KK,BN3} proved 
that a non-compact U(1) Higgs model in Landau gauge shows $U(1)$ symmetry breaking in dimensions $D\ge 3$. 
Since the classical scalar order parameter is constructed from the Cooper pair amplitudes in GL theory, 
the result by Kennedy and King encourages physicists who study a Higgs phenomenon in superconductivity.  
In other general $\alpha$-gauges \cite{KK,BN1,BN2}, however, the two-point correlations do not exhibit long-range order 
for the same Higgs model in dimensions $D\le 4$.
In addition, microscopic constitutions of superconducting materials are nothing but electrons, 
which are fermionic particles, and hence effectiveness of a gauge fixing is questionable 
for realizing a nonvanishing Cooper pair amplitude for a superconductor. 
Namely, for interacting electrons in dynamical electromagnetic fields with a specific gauge fixing,
it is highly non-trivial whether the Cooper pair amplitude can be nonvanishing as in BCS theory.

In this paper, as a concrete model of superconductor coupled to electromagnetic fields, 
we study attractively interacting lattice fermions in U(1) gauge fields. 
We treat classical compact U(1) gauge fields and quantum noncompact U(1) gauge fields. 
A gauge-fixing term which is often called $\alpha$-gauge is introduced into both of the Hamiltonians.  
By using a complex phase method \cite{McBSp,GJ,DF,Ito,KK,BN1,KomaTasaki}, we prove upper bounds 
for the two-point Cooper pair correlation at finite temperatures in spatial dimensions $D\le 4$. 
In particular, the upper bounds decay exponentially in three dimensions, and by power law in four dimensions.
These imply absence of the long-range order for the Cooper pair amplitude as a consequence of fluctuations of the gauge fields. 
Our results cannot be obtained as a corollary of Elitzur's theorem because the Hamiltonians contain the gauge-fixing term. 

The present paper is organized as follows: In the next section, we define our models and 
state our main results, Theorems~\ref{thm:classical} and \ref{thm:quantum}. 
In Sec.~\ref{CLimitGF}, we discuss a continuum limit of the gauge fields. 
The proofs of Theorems~\ref{thm:classical} and \ref{thm:quantum} are given in 
Sec.~\ref{ProofClassical} and Sec.~\ref{Proofquantum}, respectively. 
Appendices~\ref{Sec:DerboundSBcoshD} and \ref{Sec:ProofLemTrPUUP} are devoted to technical estimates. 

%%%%%%%%%%%%%%%%%%%%%%%%%%%%%%%%%%%%%%%%%%%
\Section{Models and Main Results}

\subsection{Classical Gauge Fields}

Consider first lattice fermions coupled to classical compact U(1) gauge fields. 
The tight-binding model for electrons with spin-1/2 is defined on 
a $D$-dimensional finite hypercubic lattice $\Lambda$ which is given by  
$$
\Lambda:=\{-L,-L+1,\ldots,-1,0,1,\ldots,L-1,L\}^D\subset\ze^D
$$
for $D\le 4$, and with a large positive integer $L$. In order to take into account the fluctuations of an electromagnetic field, 
we introduce a classical $U(1)$ gauge field $A$.  
The Hamiltonian $H_\Lambda(A)$ for electrons is given by 
\begin{eqnarray}
H_\Lambda(A)&:=&\sum_{x,y\in\Lambda}\sum_{\alpha,\beta}t_{x,y}^{\alpha,\beta}e^{iA_{x,y}}
c_{x,\alpha}^\dagger c_{y,\beta}\nonumber \\ 
&+&\sum_{I\ge 1}\sum_{x_1,\sigma_1}\sum_{x_2,\sigma_2}\cdots\sum_{x_I,\sigma_I}
W_{x_1,\sigma_1;x_2,\sigma_2;\ldots;x_I,\sigma_I}n_{x_1,\sigma_1}n_{x_2,\sigma_2}\cdots n_{x_I,\sigma_I},
\label{HamLambda}
\end{eqnarray}
where $c_{x,\sigma}^\dagger,c_{x,\sigma}$ are, respectively, the creation and annihilation electron operators 
at the site $x\in\Lambda$ with spin $\sigma=\uparrow,\downarrow$; 
the hopping amplitudes $t_{x,y}^{\alpha,\beta}$ are complex numbers which satisfy 
the Hermitian conditions, 
$$
t_{y,x}^{\beta,\alpha}=\left(t_{x,y}^{\alpha,\beta}\right)^\ast,
$$
and the coupling constants $W_{x_1,\sigma_1;x_2,\sigma_2;\ldots;x_I,\sigma_I}$ of the interactions are 
real numbers. As usual, we have written $n_{x,\sigma}=c_{x,\sigma}^\dagger c_{x,\sigma}$ 
for the number operators of the electron with spin $\sigma$ at the site $x$. We assume that 
the Hamiltonian $H_\Lambda(A)$ contains only the nearest neighbor hopping, 
and that the interactions are of finite range. We assume that all of the strengths are uniformly bounded as 
$$
\left|t_{y,x}^{\beta,\alpha}\right|\le t_0 \quad\mbox{and}\quad 
\left|W_{x_1,\sigma_1;x_2,\sigma_2;\ldots;x_I,\sigma_I}\right|\le W_0 
$$
with some positive constants, $t_0$ and $W_0$. 
For each nearest neighbor pair $\langle x,y\rangle$ of 
sites $x,y\in\Lambda$, 
the gauge field $A_{x,y}$ takes the value $A_{x,y}\in\re\ \mbox{mod}\ 2\pi$, and satisfies the conditions, 
$$
A_{y,x}=-A_{x,y}\ \mbox{mod}\ 2\pi.
$$

The total energy which contains the energy of the gauge fields and the gauge fixing term, is  
\begin{equation}
\label{energy}
\mathcal{H}_\Lambda(A):=H_\Lambda(A)-\mu N_\Lambda-\lambda\sum_p\cos B_p
-\frac{1}{\alpha}\sum_x\cos(d^\ast A)_x,
\end{equation}
where $N_\Lambda$ is the total number operator of electrons with the chemical potential $\mu$, i.e., 
$$
N_\Lambda:=\sum_{\sigma}\sum_{x\in\Lambda}n_{x,\sigma},
$$
and $B_p$ is the magnetic flux through the plaquette $p$ (unit square cell).  
The parameter $\lambda$ is taken to be positive. The last term is the gauge fixing term \cite{BN1} 
with the gauge parameter $\alpha>0$. The divergence $d^\ast A$ of the gauge field $A$ is given by 
$$
(d^\ast A)_x=\sum_{i=1}^D(A_{x+e_i,x}-A_{x,x-e_i}),
$$
where $e_i$ is the unit vector in the $i$-th direction. In order to avoid the appearance of the remaining 
gauge degree of freedom which is called the Gribov ambiguity,   
we impose the open boundary conditions for the hypercubic lattice $\Lambda$. 
Therefore, we set $A_{x,y}=0$ for $x\notin\Lambda$. 
The Gribov ambiguity trivially yields absence of the $U(1)$ symmetry breaking \cite{BN1}.

We consider the Cooper pair $c_{u,\uparrow}c_{v,\downarrow}$ of electrons for fixed two sites $u,v$.
We choose $v=u+a$ with a constant vector $a$.  
The two-point correlation function is given by 
$$
\left\langle c_{u,\uparrow}^\dagger c_{v,\downarrow}^\dagger c_{v',\downarrow}c_{u',\uparrow}\right\rangle_\Lambda,
$$
where $v'=u'+a$, and the expectation value at an inverse temperature $\beta$ is given by 
\begin{equation}
\label{ExpValu}
\langle \cdots\rangle_\Lambda:=
\frac{1}{Z_\Lambda}\int_{-\pi}^{\pi}\prod_{b\in\mathcal{B}}dA_b\; 
{\rm Tr}\; (\cdots)\exp[-\beta \mathcal{H}_\Lambda(A)], 
\end{equation}
where $Z_\Lambda$ is the partition function, and $\mathcal{B}$ is the set of the bonds 
(the nearest neighbor pairs of sites). 
Although all the magnetic flux $B_p$ through the plaquette $p$ are 
vanishing in the infinite limit $\lambda\uparrow\infty$, the gauge fixing degree of freedom still remains 
in the limit. For the expectation value in the infinite-volume limit, we write 
$$
\left\langle c_{u,\uparrow}^\dagger c_{v,\downarrow}^\dagger c_{v',\downarrow}c_{u',\uparrow}\right\rangle
=\lim_{\Lambda\nearrow\ze^D}
\left\langle c_{u,\uparrow}^\dagger c_{v,\downarrow}^\dagger c_{v',\downarrow}c_{u',\uparrow}\right\rangle_\Lambda.
$$
By using a complex phase method \cite{McBSp,GJ,DF,Ito,KK,BN1,KomaTasaki}, we prove: 

\begin{theorem} 
\label{thm:classical}
{\rm ({\bf Classical compact U(1) gauge fields})} In three and four dimensions, $D=3,4$, the two-point Cooper pair correlation 
function satisfies upper bounds,  
\begin{eqnarray}
\left|\left\langle c_{u,\uparrow}^\dagger c_{v,\downarrow}^\dagger c_{v',\downarrow}c_{u',\uparrow}\right\rangle\right|
&\le& {\rm Const.}\times\cases{\exp\left[-{\cal C}_4f(\beta/\alpha) \log|u-u'| \right], & if $D=4$;\cr 
\exp\left[-{\cal C}_3f(\beta/\alpha) |u-u'| \right], & if $D=3$, \cr}
\end{eqnarray}
where ${\cal C}_D$ is a positive constant which depends on the dimension $D$, and 
$$
f(\beta/\alpha):=16\sup_{q>0}\left\{\kappa q-\frac{\beta}{(c_0)^2\alpha}\frac{\cosh qJ-1}{J^2}\right\}
$$
with positive constants, $\kappa$, $c_0$ and $J$.  For a large $\beta/\alpha$, the function $f$ behaves as  
$$
f(\beta/\alpha)\sim \alpha/\beta. 
$$
\end{theorem}
The proof is given in Sec.~\ref{ProofClassical}. 
\bigskip

\noindent
{\it Remark:} 
(i) For two dimensions, we can prove an exponentially decaying upper bound for the Cooper pair correlation 
in the same way. See the remark at the end of Sec.~\ref{ProofClassical}. 
\medskip

\noindent
(ii) In the limit $\alpha \searrow 0$, the bounds in Theorem~\ref{thm:classical} are trivial, i.e., 
we cannot obtain any information about the decay of the correlation. 
Similar situations already occur in Higgs models. Namely, except for Landau gauge, other gauge fixings show 
absence of symmetry breaking \cite{FMS,KK,BN1,BN2}. In the situation in the continuum limit of 
the gauge fields, however, it may change because we can take the limit $\alpha \searrow 0$ simultaneously 
with the continuum limit of the gauge fields. We discuss the continuum limit in Sec.~\ref{CLimitGF} below. 
\medskip

\noindent
(iii) From our results, we cannot make a definite conclusion about the Higgs mechanism, 
which leads to photon mass generation, i.e., Meissner effect in superconductors. 
Besides, we cannot elucidate whether or not the power-law decaying upper bound for the correlation in four dimensions 
has a physical meaning. These issues are left for future studies.
 
%%%%%%%%%%%%%%%%%%%%%%%%%%%%%%%%%%%%%%%%%%%%%%%%%%
\subsection{Quantum Gauge Fields}

Consider a noncompact $U(1)$ gauge fields $A$. Namely, for each nearest neighbor pair $\langle x,y\rangle$ of 
sites $x,y\in\Lambda$, the gauge field $A_{x,y}$ takes the value $A_{x,y}\in\re$. 
The Hamiltonian of the quantized gauge field is given by 
\begin{equation}
\label{hamquantumGF}
H_\Lambda^{\rm g}:=\frac{1}{2g_{\rm e}}\sum_{\langle x,y\rangle}E_{x,y}^2
+\frac{1}{2g_{\rm m}}\sum_p B_p^2+\frac{1}{\alpha}\sum_{x\in\Lambda}(d^\ast A)_x^2,
\end{equation}
where $g_{\rm e}$ and $g_{\rm m}$ are positive coupling constants, and 
the electric field $E_{x,y}$ is the canonical conjugate momentum \cite{Feynman} for the gauge field $A_{x,y}$, i.e.,   
$$
E_{x,y}:=-i\frac{\partial}{\partial A_{x,y}}.
$$ 
In other words, the gauge fields can be interpreted as a quantum coupled oscillators. 
We impose the same open boundary condition as that of \cite{KK} so that 
there appears no zero mode of the gauge fields. 
Hence, the expectation value (\ref{expectNonCompact}) below is well defined.
In passing, we can also 
impose an alternative boundary condition which does not yield zero modes of the gauge fields 
as in \cite{BN3}.

The total energy is given by 
\begin{equation}
\label{TotenergyQuantum}
\mathcal{H}_\Lambda:=H_\Lambda(A)-\mu N_\Lambda+H_\Lambda^{\rm g}.
\end{equation}
The expectation value is given by 
\begin{equation}
\label{expectNonCompact}
\langle \cdots \rangle_\Lambda:=\frac{1}{Z_\Lambda}{\rm Tr}\;(\cdots)\exp[-\beta\mathcal{H}_\Lambda]. 
\end{equation}
For the expectation value in the infinite-volume limit, we write 
$$
\left\langle c_{u,\uparrow}^\dagger c_{v,\downarrow}^\dagger c_{v',\downarrow}c_{u',\uparrow}\right\rangle
=\lim_{\Lambda\nearrow\ze^D}
\left\langle c_{u,\uparrow}^\dagger c_{v,\downarrow}^\dagger c_{v',\downarrow}c_{u',\uparrow}\right\rangle_\Lambda.
$$ 
By using a complex phase method \cite{McBSp,GJ,DF,Ito,KK,BN1,KomaTasaki}, we obtain:

\begin{theorem}
\label{thm:quantum}
{\rm ({\bf Quantum noncompact U(1) gauge fields})} In three and four dimensions, $D=3,4$, 
the two-point Cooper pair correlation function satisfies upper bounds, 
\begin{eqnarray}
\left|\left\langle c_{u,\uparrow}^\dagger c_{v,\downarrow}^\dagger c_{v',\downarrow}c_{u',\uparrow}\right\rangle\right|
&\le& {\rm Const.}\times\cases{\exp\left[-\tilde{\cal C}_4\alpha\beta^{-1} \log|u-u'| \right], & if $D=4$;\cr 
\exp\left[-\tilde{\cal C}_3\alpha\beta^{-1} |u-u'| \right], & if $D=3$, \cr}
\end{eqnarray}
where $\tilde{\cal C}_3$ and $\tilde{\cal C}_4$ are some positive constants. 
\end{theorem}

The proof is given in Sec.~\ref{Proofquantum}. 
\bigskip

\noindent
{\it Remark:} (i) In this case, the statements in Remarks of Theorem~\ref{thm:classical} holds, too.  
In particular, the limit $\alpha\searrow 0$ gives the well known Coulomb gauge.

%%%%%%%%%%%%%%%%%%%%%%%%%%%%%%%%%%%%%%%%%%%%%%%%%%%%%
\Section{Continuum Limit of the Gauge Fields}
\label{CLimitGF}

In this section, we consider a continuum limit of the standard electromagnetic fields in three spatial dimensions.
The Hamiltonian is given by (\ref{hamquantumGF}) in the preceding section. 

For the purpose of the present section, we consider a cubic lattice $\ell\ze^3$ with the lattice constant $\ell$ 
which is a large positive integer. 
We embed this lattice $\ell\ze^3$ into the cubic lattice $\ze^3$ with unit length. 
The tight-binding model for electrons is defined on the lattice $\ell\ze^3$. 
This implies that the scale of the interatomic distance of the crystal is given by $\ell$. 
On the other hand, the electromagnetic fields are defined on the lattice $\ze^3$. 
Therefore, each hopping term in the Hamiltonian of the tight-binding model is replaced with 
$$
t_{x,y}^{\alpha,\beta}c_{x,\alpha}^\dagger c_{y,\beta}\exp\Bigl[i\sum_{i=0}^{\ell-1} A_{x_i,x_{i+1}}\Bigr]
$$
for $x,y\in \ell\ze^3$, $(x\ne y)$, where the $\ell+1$ lattice sites, $x_0,x_1,\ldots, x_\ell\in \ze^3$, satisfy 
the following conditions: $x_0=x$, $x_\ell=y$, and the bonds, $\langle x_i,x_{i+1}\rangle $, are neighboring two sites, i.e., 
these bonds form the straight path which connects $x$ and $y$. 
 
Now, we consider the continuum limit $\ell\nearrow\infty$. We recall our bound for the correlation as 
\begin{equation}
\label{CCScaling}
\left|\left\langle c_{u,\uparrow}^\dagger c_{v,\downarrow}^\dagger c_{v',\downarrow}c_{u',\uparrow}\right\rangle\right|
\le{\rm Const.}\exp\left[-\tilde{\cal C}_3\alpha\beta^{-1} |u-u'| \right]. 
\end{equation}
We introduce the distance in the crystal as $r=|u-u'|/\ell$. Then, the upper bound is written as 
$$
\exp\left[-\tilde{\cal C}_3\alpha\beta^{-1} \ell r \right]. 
$$
In order to realize massless photons \cite{Guth,FroeSpen}, 
we have to take a scaling limit for the coupling constants, $g_{\rm e}$, $g_{\rm m}$ and $\alpha$. 
Since the expectation value (\ref{CCScaling}) is bounded with respect to all the parameters, 
the corresponding limit for the correlation obviously exists by taking a suitable subsequence. However, the existence of 
the continuum limit of the Hamiltonian of the gauge fields is unclear. Since this problem is beyond the 
scope of the present paper, we consider only the Cooper pair correlation.  

Consider first the case that the coupling constant $\alpha$ goes to some positive $\alpha_{\rm C}$ 
in the continuum limit $\ell\nearrow\infty$. In this case, the upper bound vanishes in the limit. 
Therefore, the correlation is vanishing for any distance $r$ of the crystal. 

Next, consider the case that $\alpha$ goes to zero in the limit. 
In this case, we can fix the value of $\alpha \ell$ to some finite non-zero value, 
and then we can take the limit $\ell\nearrow\infty$. 
In consequence, the upper bound still exponentially decays in distance $r$ of the crystal in the limit $\ell\nearrow\infty$.

%%%%%%%%%%%%%%%%%%%%%%%%%%%%%%%%%%%%%%%%%%%%%%%%%%%%%%%%%%%%
\Section{Proof of Theorem~\ref{thm:classical}}
\label{ProofClassical} 

Following Koma and Tasaki \cite{KomaTasaki}, we introduce a complex gauge transformation $\Gamma(\varphi)$ 
with a real function $\varphi_x$ on the site $x\in\Lambda$ as 
$$
\Gamma(\varphi):=\prod_{x,\sigma}e^{-\varphi_x n_{x,\sigma}}.
$$
Then, one has   
$$
\Gamma(\varphi) c_{x,\sigma}^\dagger\Gamma(\varphi)^{-1}=e^{-\varphi_x}c_{x,\sigma}^\dagger 
$$
and 
$$
\Gamma(\varphi) c_{x,\sigma}\Gamma(\varphi)^{-1}=e^{\varphi_x}c_{x,\sigma}.
$$
Further, we introduce the corresponding gauge transformation \cite{McBSp} for the gauge field $A$ as 
\begin{equation}
\label{Atrans}
A\rightarrow A-id\varphi,
\end{equation}
where the gradient $d\varphi$ of the function $\varphi$ is given by 
$$
(d\varphi)_{x,y}=\varphi_x-\varphi_y
$$
for the nearest neighbor pair $x,y$ of the sites. This transformation is realized by using 
a contour integral with respect to the gauge field $A$ along the rectangular path in the complex plane 
in the right-hand side of (\ref{ExpValu}). Then, the lateral contours cancel each other out 
due to the periodicity of the cosine. 

The hopping terms in the Hamiltonian ${H}_\Lambda(A)$ are invariant under the transformation $\Gamma(\varphi)$ as  
\begin{eqnarray*}
c_{x,\alpha}^\dagger c_{y,\beta}e^{iA_{xy}}\rightarrow 
\Gamma(\varphi)c_{x,\alpha}^\dagger c_{y,\beta}\Gamma(\varphi)^{-1}e^{i[A_{xy}-i(\varphi_x-\varphi_y)]}&=&
e^{-\varphi_x}c_{x,\alpha}^\dagger e^{\varphi_y}c_{y,\beta}e^{iA_{xy}}e^{\varphi_x-\varphi_y}\\
&=&c_{x,\alpha}^\dagger c_{y,\beta}e^{iA_{xy}}. 
\end{eqnarray*}
Similarly, the magnetic flux $B_p$ is also invariant under the transformation (\ref{Atrans}). 
The divergence of the gauge field $A$ is transformed as 
$$
d^\ast A\rightarrow d^\ast A-i\Delta\varphi.
$$
where we have written $\Delta=d^\ast d$ in terms of the Laplacian $\Delta$. 

By using the transformation and the property of the trace, we have 
\begin{eqnarray*}
& &\int_{-\pi}^{\pi}\prod_{b\in\mathcal{B}}dA_b\; 
{\rm Tr}\;
c_{u,\uparrow}^\dagger c_{v,\downarrow}^\dagger c_{v',\downarrow}c_{u',\uparrow}
\exp[-\beta \mathcal{H}_\Lambda(A)]\\
&=&\int_{-\pi}^{\pi}\prod_{b\in\mathcal{B}}dA_b\; 
{\rm Tr}\;\Gamma(\varphi)
c_{u,\uparrow}^\dagger c_{v,\downarrow}^\dagger c_{v',\downarrow}c_{u',\uparrow}\Gamma(\varphi)^{-1}
\exp[-\beta \Gamma(\varphi)\mathcal{H}_\Lambda(A)\Gamma(\varphi)^{-1}]\\
&=&e^{-\varphi_u-\varphi_v}e^{\varphi_{u'}+\varphi_{v'}}
\int_{-\pi}^{\pi}\prod_{b\in\mathcal{B}}dA_b\; 
{\rm Tr}\;
c_{u,\uparrow}^\dagger c_{v,\downarrow}^\dagger c_{v',\downarrow}c_{u',\uparrow}
\exp[-\beta \mathcal{H}_\Lambda(A)-\beta\delta\mathcal{H}_\Lambda(A)],
\end{eqnarray*}
where 
$$
\delta\mathcal{H}_\Lambda(A)=-\frac{1}{\alpha}\sum_x\left\{\cos(d^\ast A)_x[\cosh(\Delta \varphi)_x-1]
-i\sin(d^\ast A)_x\sinh(\Delta\varphi)_x\right\},
$$
and we have used  
$$
\cos(d^\ast A-i\Delta \varphi)_x=\cos(d^\ast A)_x\cosh(\Delta \varphi)_x -i\sin(d^\ast A)_x\sinh(\Delta\varphi)_x. 
$$
Further, by using the bound, 
$$
\cos(d^\ast A)_x[\cosh(\Delta \varphi)_x-1]
\le \cosh(\Delta \varphi)_x-1,
$$
we obtain 
\begin{eqnarray*}
& &\left|\int_{-\pi}^{\pi}\prod_{b\in\mathcal{B}}dA_b\; 
{\rm Tr}\;
c_{u,\uparrow}^\dagger c_{v,\downarrow}^\dagger c_{v',\downarrow}c_{u',\uparrow}
\exp[-\beta \mathcal{H}_\Lambda(A)]\right|\\
&\le& 
Z_\Lambda e^{-\varphi_u-\varphi_v}e^{\varphi_{u'}+\varphi_{v'}}
\exp\Bigl[\frac{\beta}{\alpha}\sum_x\{\cosh(\Delta \varphi)_x-1\}\Bigr]. 
\end{eqnarray*}
This implies 
\begin{equation}
\label{CooperTwistBound}
\left|\Bigl\langle c_{u,\uparrow}^\dagger c_{v,\downarrow}^\dagger 
c_{v',\downarrow}c_{u',\uparrow}\Bigr\rangle_\Lambda\right|
\le e^{-\varphi_u-\varphi_v}e^{\varphi_{u'}+\varphi_{v'}}
\exp\Bigl[\frac{\beta}{\alpha}\sum_x\{\cosh(\Delta \varphi)_x-1\}\Bigr].
\end{equation}

In order to estimate the right-hand side of (\ref{CooperTwistBound}), 
we introduce a cutoff set of wavenumbers as  
$$
\mathcal{M}_\epsilon^K:=\{k\; | \; \epsilon\le |k^{(i)}|\le K, \ i=1,2,\ldots,D \},
$$
where the wavenumber $k$ is given by 
$$
k=(k^{(1)},k^{(2)},\ldots,k^{(D)})\in\re^D
$$
with $k^{(i)}=2\pi n^{(i)}/(2L+1)$ with the integer $n^{(i)}\in\{-L,-L+1,\ldots,-1,0,1,\ldots,L\}$, 
and the cutoff parameters, $\epsilon$ and $K$, are positive.   
We choose the cutoff $K$ to satisfy the following two conditions: 
\begin{equation}
\label{condition1}
1-\cos k^{(i)}\ge c_0(k^{(i)})^2\quad \mbox{for \ } k^{(i)}\in[-K,K]
\end{equation}
and 
\begin{equation}
\label{condition2}
1+\cos(k\cdot a)\ge \kappa \quad \mbox{for \ } k\in \mathcal{M}_\epsilon^K, 
\end{equation}
where both of $c_0$ and $\kappa$ are a positive constant. 

Consider first the case of $D=3,4$. 
By using the cutoff set $\mathcal{M}_\epsilon^K$, we choose the function $\varphi$ as 
$$
\varphi_x=\frac{q}{|\Lambda|}\sum_y\sum_{k\in\mathcal{M}_\epsilon^K}
\frac{e^{ik\cdot(x-y)}p_y}{\Bigl[\sum_{i=1}^D(1-\cos k^{(i)})\Bigr]^2}
$$
with a ``charge" $q\in\re$ which will be determined later, where 
$p_y=\delta_{y,u}-\delta_{y,u'}$ with the Kronecker delta $\delta_{x,y}$. 
We note that 
\begin{equation}
\label{varphix}
\varphi_x=-\frac{2iq}{|\Lambda|}%\sum_y
\sum_{k\in\mathcal{M}_\epsilon^K}
\frac{e^{ik\cdot x}e^{-ik\cdot(u+u')/2}}{\Bigl[\sum_{i=1}^D(1-\cos k^{(i)})\Bigr]^2}\sin[k\cdot(u-u')/2].
\end{equation}
Since the function $\varphi$ clearly has the periodicity with respect to the spatial coordinates, 
$\Delta\varphi$ is not necessarily compatible with the present open boundary condition. Therefore, we have to show 
that these differences at the boundaries are negligible in the infinite-volume limit. 

{From} the definition of the Laplacian $\Delta$, one has 
$$
(\Delta\varphi)_x=\sum_{i=1}^D(\varphi_{x+e_i}+\varphi_{x-e_i}-2\varphi_x) 
$$
for $x\in \Lambda\backslash \partial\Lambda$, i.e., the interior of $\Lambda$, 
where $\partial\Lambda$ is the boundary of $\Lambda$. When $x\in\partial\Lambda$, e.g., 
$x=(L,x^{(2)},\ldots,x^{(D)})$, the first term is different from the above as 
$$
(\Delta\varphi)_x=\varphi_{x-e_1}-\varphi_x+\sum_{i=2}^D(\varphi_{x+e_i}+\varphi_{x-e_i}-2\varphi_x). 
$$
This can be rewritten as 
\begin{equation}
\label{Deltavarphi}
(\Delta\varphi)_x=(\tilde{\Delta}\varphi)_x-(\varphi_{x+e_1}-\varphi_x),
\end{equation}
where the extended Laplacian $\tilde{\Delta}$ is defined by 
$$
(\tilde{\Delta}\psi)_x=\sum_{i=1}^D(\psi_{x+e_i}+\psi_{x-e_i}-2\psi_x)
\quad \mbox{for all \ } x\in\Lambda
$$
and for a function $\psi$ on $\Lambda$. By definition, one has 
\begin{equation}
\label{tildeDeltavarphi}
(\tilde{\Delta}\varphi)_x=-\frac{4i q}{|\Lambda|}\sum_{k\in\mathcal{M}_\epsilon^K}
\frac{e^{ik\cdot x}e^{-ik\cdot(u+u')/2}}{\sum_{i=1}^D(1-\cos k^{(i)})}\sin[k\cdot(u-u')/2].
\end{equation}
{From} this and the assumption $D\ge 3$, one can show 
\begin{equation}
\label{LaplacevarphiBound}
|(\tilde{\Delta}\varphi)_x|\le qJ,
\end{equation} 
where $J$ is a positive constant which is independent of $\Lambda$, $x$, $u$, $u'$ and the cutoff parameter $\epsilon$. 

Now, let us estimate the right-hand side of (\ref{CooperTwistBound}). 
Note that 
$$
\sum_{x\in\Lambda}\{\cosh(\Delta \varphi)_x-1\}
=\sum_{x\in\Lambda\backslash\partial\Lambda}\{\cosh(\tilde{\Delta}\varphi)_x-1\}
+\sum_{x\in\partial\Lambda}\{\cosh(\Delta \varphi)_x-1\}.
$$
Using the bound (\ref{LaplacevarphiBound}), the first sum in the right-hand side can be evaluated as 
$$
\sum_{x\in\Lambda\backslash\partial\Lambda}\{\cosh(\tilde{\Delta}\varphi)_x-1\}
\le \frac{\cosh qJ-1}{(qJ)^2}\sum_{x\in\Lambda\backslash\partial\Lambda}|\tilde{\Delta}\varphi)_x|^2.
$$
The second sum can be estimated as 
\begin{equation}
\label{SumBoundarycoshDeltaPhi}
\sum_{x\in\partial\Lambda}[\cosh(\Delta\varphi)_x-1]\le \frac{\cosh qJ-1}{(qJ)^2}\sum_{x\in\partial\Lambda}
|(\tilde{\Delta}\varphi)_x|^2+\delta C,
\end{equation}
where $\delta C$ is the correction which goes to zero in the limit $L\rightarrow\infty$. 
The derivation of (\ref{SumBoundarycoshDeltaPhi}) is given in Appendix~\ref{Sec:DerboundSBcoshD}. 
Combining these with (\ref{tildeDeltavarphi}), we obtain  
\begin{eqnarray}
\sum_{x\in\Lambda}[\cosh(\Delta\varphi)_x-1]&\le& \frac{\cosh qJ-1}{(qJ)^2}\sum_{x\in\Lambda}
|(\tilde{\Delta}\varphi)_x|^2+\delta C \nonumber\\
&=&\frac{16(\cosh qJ-1)}{J^2}\frac{1}{|\Lambda|}\sum_{k\in\mathcal{M}_\epsilon^K}
\frac{\sin^2[k\cdot (u-u')/2]}{\Bigl[\sum_{i=1}^D(1-\cos k^{(i)})\Bigr]^2}+\delta C.
\label{coshDeltavarphiest}
\end{eqnarray}
In order to estimate the sum in the right-hand side, we write 
$$
g_\epsilon(u-u'):=\frac{1}{(2\pi)^D}\int_{k\in\mathcal{M}_\epsilon^K}dk^{(1)}\cdots dk^{(D)}
\frac{\sin^2 [k\cdot (u-u')/2]}{[\sum_{i=1}^D(1-\cos k^{(i)})]^2}
$$
in the infinite-volume limit $\Lambda\nearrow\ze^D$. We also write 
$$
g(u-u'):=\lim_{\epsilon \searrow 0}g_\epsilon(u-u'). 
$$
By using the condition (\ref{condition1}) for the cutoff $K$, we have 
$$
g(u-u')\le\frac{1}{(2\pi )^D(c_0)^2}
\int d\Omega_D\int_0^K dr\; \frac{|\sin^2(|u-u'||\cos \theta|r/2)|}{r^{5-D}},
$$
where we have introduced the polar coordinates, $r=|k|$ and $\theta$, and $\Omega_D$ is the solid angle.
By changing the variable as $t=|u-u'||\cos\theta|r/2$, the integral is written 
\begin{eqnarray}
\frac{1}{(2\pi)^D}\int_0^K dr\; \frac{\sin^2(|u-u'||\cos \theta|r/2)}{r^{5-D}}
&=&\frac{(|u-u'||\cos\theta|/2)^{4-D}}{(2\pi)^D}\int_0^{|u-u'||\cos\theta|K/2} dt\; \frac{\sin^2 t}{t^{5-D}}\nonumber \\
&\sim& {\cal C}_D\times \cases{\log|u-u'|, & if $D=4$;\cr 
|u-u'|, & if $D=3$, \cr}
\label{Intest}
\end{eqnarray}
where ${\cal C}_D$ is the positive constant, which depends on the dimension $D$. 
Substituting this into the above right-hand side, we obtain 
\begin{equation}
\label{gestimate}
g(u-u')\le\frac{{\cal C}_D}{(c_0)^2}\times\cases{\log|u-u'|, & if $D=4$;\cr 
|u-u'|, & if $D=3$. \cr}
\end{equation}

Next, consider the quantity $\varphi_u-\varphi_{u'}+\varphi_v-\varphi_{v'}$ 
in the right-hand side of (\ref{CooperTwistBound}). From the expression (\ref{varphix}) of $\varphi_x$, 
we have  
$$
\varphi_u-\varphi_{u'}=\frac{4q}{|\Lambda|}\sum_{k\in\mathcal{M}_\epsilon^K}
\frac{\sin^2[ k\cdot( u-u')/2]}{[\sum_{i=1}^D(1-\cos k^{(i)})]^2},
$$
and
$$
\varphi_v-\varphi_{v'}=\frac{4q}{|\Lambda|}\sum_{k\in\mathcal{M}_\epsilon^K}
\frac{\cos(k\cdot a)\sin^2[ k\cdot (u-u')/2]}{[\sum_{i=1}^D(1-\cos k^{(i)})]^2},
$$
where we have used $v=u+a$ and $v'=u'+a$. By adding both sides, one has  
$$
\varphi_u-\varphi_{u'}+\varphi_v-\varphi_{v'}=
\frac{4q}{|\Lambda|}\sum_{k\in\mathcal{M}_\epsilon^K}
\frac{[1+\cos(k\cdot a)]\sin^2[ k\cdot (u-u')/2]}{[\sum_{i=1}^D(1-\cos k^{(i)})]^2}.
$$
In the infinite-volume limit $\Lambda\nearrow\ze^D$, one has 
$$
\varphi_u-\varphi_{u'}+\varphi_v-\varphi_{v'} 
=\frac{4q}{(2\pi)^D}\int_{k\in\mathcal{M}_\epsilon^K}dk^{(1)}\cdots dk^{(D)}
\frac{[1+\cos(k\cdot a)]\sin^2 [k\cdot (u-u')/2]}{[\sum_{i=1}^D(1-\cos k^{(i)})]^2}.
$$
We write 
$$
\tilde{g}_\epsilon(u-u'):=\frac{4}{(2\pi)^D}\int_{k\in\mathcal{M}_\epsilon^K}dk^{(1)}\cdots dk^{(D)}
\frac{[1+\cos(k\cdot a)]\sin^2 [k\cdot (u-u')/2]}{[\sum_{i=1}^D(1-\cos k^{(i)})]^2},
$$
and 
$$
\tilde{g}(u-u'):=\lim_{\epsilon \searrow 0}\tilde{g}_\epsilon(u-u').
$$
By using $1-\cos k^{(i)}\le (k^{(i)})^2/2$ and the condition (\ref{condition2}) for the cutoff $K$, 
the integral of $\tilde{g}(u-u')$ can be estimated as  
\begin{eqnarray*}
& &\frac{1}{(2\pi)^D}\int dk^{(1)}\cdots dk^{(D)}\; \frac{[1+\cos(k\cdot a)]\sin^2 k\cdot (u-u')/2}{[\sum_{i=1}^D(1-\cos k^{(i)})]^2}\\
&\ge& \frac{4}{(2\pi)^D}\int dk^{(1)}\cdots dk^{(D)}\; \frac{[1+\cos(k\cdot a)]\sin^2 k\cdot (u-u')/2}{|k|^4}\\
&\ge &\frac{4\kappa}{(2\pi)^D} \int d\Omega_D\int_0^{K} dr\; \frac{\sin^2(|u-u'||\cos \theta|r/2)}{r^{5-D}}
\end{eqnarray*}
in the same way as in the above argument. Further, the integral in the right-hand side can be 
estimated by using (\ref{Intest}). As a result, we obtain 
\begin{equation}
\varphi_u+\varphi_v-\varphi_{u'}-\varphi_{v'}
\ge{\rm Const.} + {16 {\cal C}_D}\kappa q\times\cases{\log|u-u'|, & if $D=4$;\cr 
|u-u'|, & if $D=3$. \cr}
\end{equation}
Combining this, (\ref{gestimate}), (\ref{coshDeltavarphiest}) and (\ref{CooperTwistBound}), 
the Cooper pair correlation in the infinite-volume limit can be estimated as 
\begin{eqnarray}
\left|\left\langle c_{u,\uparrow}^\dagger c_{v,\downarrow}^\dagger c_{v',\downarrow}c_{u',\uparrow}\right\rangle\right|
&\le& {\rm Const.}\times\cases{\exp\left[-{\cal C}_4f(\beta/\alpha) \log|u-u'| \right], & if $D=4$;\cr 
\exp\left[-{\cal C}_3f(\beta/\alpha) |u-u'| \right], & if $D=3$, \cr}
\end{eqnarray}
where 
$$
f(\beta/\alpha):=16\sup_{q>0}\left\{\kappa q-\frac{\beta}{(c_0)^2\alpha}\frac{\cosh qJ-1}{J^2}\right\}.
$$
Here, for a large $\beta/\alpha$, one has 
$$
f(\beta/\alpha)\sim \alpha/\beta. 
$$
\medskip

\noindent
{\it Remark:} 
For the case of two dimensions, $D=2$, we choose the function $\varphi$ as 
$$
\varphi_x=\frac{q}{|\Lambda|}\sum_y\sum_{k\in\mathcal{M}_\epsilon^K}
\frac{e^{ik\cdot(x-y)}p_y}{\Bigl[\sum_{i=1}^D(1-\cos k^{(i)})\Bigr]^{3/2}}. 
$$
In the same way, we obtain an exponential decay bound 
for the two-point Cooper pair correlation. 

%%%%%%%%%%%%%%%%%%%%%%%%%%%%%%%%%%%%%%%%%%%%%%%%%%%%%%%%%%%%%%%%%%%%%
\Section{Proof of Theorem~\ref{thm:quantum}}
\label{Proofquantum}

In order to use the Trotter formula \cite{RSI}, we introduce a mass term into 
the Hamiltonian $\mathcal{H}_\Lambda$ of (\ref{TotenergyQuantum}) as 
$$
\mathcal{H}_{\Lambda,m}:=H_\Lambda(A)-\mu N_\Lambda+H_\Lambda^{\rm g}
+\frac{m^2}{2}\sum_{\langle x,y\rangle}(A_{x,y})^2
$$
with a positive parameter $m$. After the calculations, we will take the limit $m\searrow 0$. 
One might think that this regularization is not necessary for calculation because 
there appears no zero mode \cite{KK} of the gauge field due to the gauge fixing term with the open boundary condition. 
However, this standard regularization makes the following calculation very simple without loss of mathematical rigor.
In particular, when we apply the complex phase method, there is no need to check whether or not 
there appears a new zero mode in the process of using the complex phase method.

To begin with, we decompose the Hamiltonian $\mathcal{H}_{\Lambda,m}$ into two parts as 
$$
\mathcal{H}_{\Lambda,m}=H_\Lambda^{\rm e}+\mathcal{H}_\Lambda^{\rm r},
$$
where we have written 
$$
H_\Lambda^{\rm e}:=\frac{1}{2g_{\rm e}}\sum_{\langle x,y\rangle}E_{x,y}^2,
$$
and 
$$
\mathcal{H}_\Lambda^{\rm r}:=\mathcal{H}_{\Lambda,m}-H_\Lambda^{\rm e}. 
$$
Note that 
\begin{eqnarray*}
{\rm Tr}^{\rm g}\; e^{-\beta\mathcal{H}_{\Lambda,m}}&=&\lim_{M\rightarrow\infty}
{\rm Tr}^{\rm g}\;\left(e^{-\beta H_\Lambda^{\rm e}/M}e^{-\beta\mathcal{H}_\Lambda^{\rm r}/M}\right)^M\\
&=&\lim_{M\rightarrow\infty}
{\rm Tr}^{\rm g}\;e^{-\beta H_\Lambda^{\rm e}/M}e^{-\beta\mathcal{H}_\Lambda^{\rm r}/M}\cdots
e^{-\beta H_\Lambda^{\rm e}/M}e^{-\beta\mathcal{H}_\Lambda^{\rm r}/M}, 
\end{eqnarray*}
where we have written ${\rm Tr}^{\rm g}$ for the trace with respect to the gauge fields $A$ only.

Let $\Phi(\xi)$ be a normalizable wavefunction of $\xi\in\re$. Then, one has 
$$
\Phi(\xi)=\frac{1}{2\pi}\int_{-\infty}^\infty dk\; e^{-ik\xi}\int_{-\infty}^\infty d\xi_0\; e^{ik\xi_0}\Phi(\xi_0).
$$
Let $\eta$ be a positive parameter. From the above expression, one obtains 
\begin{eqnarray*}
\exp\left[\eta\frac{d^2}{d\xi^2}\right]\Phi(\xi)&=&
\frac{1}{2\pi}\int_{-\infty}^\infty dk\; e^{-\eta k^2}e^{-ik\xi}\int_{-\infty}^\infty d\xi_0\; e^{ik\xi_0}\Phi(\xi_0)\\
&=&\frac{1}{\sqrt{4\pi\eta}}\int_{-\infty}^\infty d\xi_0\; \exp[-(\xi-\xi_0)^2/(4\eta)]\Phi(\xi_0). 
\end{eqnarray*}
By using this formula with $\eta=\beta/(2g_{\rm e}M)$, we obtain 
\begin{eqnarray*}
& &{\rm Tr}^{\rm g}\;e^{-\beta H_\Lambda^{\rm e}/M}e^{-\beta\mathcal{H}_\Lambda^{\rm r}/M}\cdots
e^{-\beta H_\Lambda^{\rm e}/M}e^{-\beta\mathcal{H}_\Lambda^{\rm r}/M}\\
&=&\int_{-\infty}^\infty \prod_{i=0}^{M-1}\prod_{\langle x,y\rangle}
\frac{dA_{x,y}^{(i)}}{\sqrt{4\pi \eta}}\; 
\exp\Bigl[-\sum_{\langle x,y\rangle}(A_{x,y}^{(M)}-A_{x,y}^{(M-1)})^2/(4\eta)\Bigr]
\exp[-\beta \mathcal{H}_\Lambda^{\rm r}(A^{(M-1)})/M]\\
& &\cdots \exp\Bigl[-\sum_{\langle x,y\rangle}(A_{x,y}^{(2)}-A_{x,y}^{(1)})^2/(4\eta)\Bigr]
\exp[-\beta \mathcal{H}_\Lambda^{\rm r}(A^{(1)})/M]\\
&\times&\exp\Bigl[-\sum_{\langle x,y\rangle}(A_{x,y}^{(1)}-A_{x,y}^{(0)})^2/(4\eta)\Bigr]
\exp[-\beta \mathcal{H}_\Lambda^{\rm r}(A^{(0)})/M],
\end{eqnarray*}
where we have written $\mathcal{H}_\Lambda^{\rm r}(A)$ for 
the Hamiltonian $\mathcal{H}_\Lambda^{\rm r}$ with the gauge fields $A$, and the gauge field $A^{(M)}$ is 
given by $A^{(M)}=A^{(0)}$.  

In the same way as in the case of the classical gauge fields, we can apply the transformation,
$$
A^{(i)}\rightarrow A^{(i)}-id\varphi,
$$
for all $i=0,1,\ldots,M-1$, and $\Gamma(\varphi)$ for the electrons. 
Clearly, one has 
$$
d^\ast A^{(i)} \rightarrow d^\ast A^{(i)} -i\Delta \varphi.   
$$
In the same way as in the case of the classical fields, we obtain   
$$
\exp[-\beta \mathcal{H}_\Lambda^{\rm r}(A^{(i)})/M]\rightarrow 
\exp[-\beta \mathcal{H}_\Lambda^{\rm r}(A^{(i)})/M]\exp[w[\varphi]/M] U,  
$$
where we have written 
\begin{equation}
\label{wvarphi}
w[\varphi]:=\frac{\beta}{\alpha}\sum_x [(\Delta\varphi)_x]^2
+\frac{\beta m^2}{2}\sum_{\langle x,y\rangle}[(d\varphi)_{x,y}]^2
\end{equation}
and
$$
U:=\exp\left[\frac{2i\beta}{M\alpha}\sum_x (d^\ast A)_x (\Delta\varphi)_x
+\frac{i\beta m^2}{M} \sum_{\langle x,y\rangle} A_{x,y}(d\varphi)_{x,y}\right]
$$
for short. Therefore, we have 
$$
\Gamma(\varphi){\rm Tr}^{\rm g}\;\left(e^{-\beta H_\Lambda^{\rm e}/M}e^{-\beta\mathcal{H}_\Lambda^{\rm r}/M}\right)^M
\Gamma(\varphi)^{-1}
=e^{w[\varphi]}\;{\rm Tr}^{\rm g}\;\left(e^{-\beta H_\Lambda^{\rm e}/M}e^{-\beta\mathcal{H}_\Lambda^{\rm r}/M}
U\right)^M.
$$
{From} these observations, one has 
\begin{eqnarray}
\left\langle c_{u,\uparrow}^\dagger c_{v,\downarrow}^\dagger c_{v',\downarrow}c_{u',\uparrow}\right\rangle_{\Lambda,m}^{(M)}
&:=&\frac{1}{Z_\Lambda}{\rm Tr}\; c_{u,\uparrow}^\dagger c_{v,\downarrow}^\dagger c_{v',\downarrow}c_{u',\uparrow}
(e^{-\beta H_\Lambda^{\rm e}/M}e^{-\beta\mathcal{H}_\Lambda^{\rm r}/M})^M \nonumber \\
&=&\frac{1}{Z_\Lambda}e^{-\varphi_u-\varphi_v+\varphi_{u'}+\varphi_{v'}}\; e^{w[\varphi]}\; 
{\rm Tr}\; c_{u,\uparrow}^\dagger c_{v,\downarrow}^\dagger c_{v',\downarrow}c_{u',\uparrow}(PU)^M,\qquad \quad 
\label{expCooperMm} 
\end{eqnarray}
where we have written 
$$
P=e^{-\beta\mathcal{H}_\Lambda^{\rm r}/(2M)}e^{-\beta H_\Lambda^{\rm e}/M}e^{-\beta\mathcal{H}_\Lambda^{\rm r}/(2M)}.
$$

We choose $M=2\ell$ with a positive integer $\ell$. Using the Schwarz inequality, we have 
\begin{equation}
\label{TrCooperPUbound}
\left|{\rm Tr}\; c_{u,\uparrow}^\dagger c_{v,\downarrow}^\dagger c_{v',\downarrow}c_{u',\uparrow}(PU)^M\right|
\le {\rm Tr}\; (PU)^\ell (U^\ast P)^\ell.
\end{equation}
The right-hand side is estimated as follows:

\begin{lemma}
\label{lem:TrPUUP}
The following bound is valid: 
\begin{equation}
\label{PUbound}
{\rm Tr}\; (PU)^\ell (U^\ast P)^\ell\le {\rm Tr}\; P^{2\ell}.
\end{equation}
\end{lemma}

The proof is given in Appendix~\ref{Sec:ProofLemTrPUUP}. 
The right-hand side of (\ref{PUbound}) converges to the partition function $Z_\Lambda$ 
in the limit $\ell\rightarrow\infty$. 
Combining this, (\ref{wvarphi}), (\ref{expCooperMm}) and (\ref{TrCooperPUbound}), we have  
\begin{eqnarray*}
\left|\left\langle c_{u,\uparrow}^\dagger c_{v,\downarrow}^\dagger 
c_{v',\downarrow}c_{u',\uparrow}\right\rangle_{\Lambda}\right|
&=& \lim_{m\searrow 0}\lim_{\ell \nearrow\infty} \left|\left\langle c_{u,\uparrow}^\dagger c_{v,\downarrow}^\dagger 
c_{v',\downarrow}c_{u',\uparrow}\right\rangle_{\Lambda,m}^{(2\ell)}\right|\\
&\le& e^{-\varphi_u-\varphi_v+\varphi_{u'}+\varphi_{v'}}
\exp\left[{\beta}\sum_x (\Delta\varphi)_x^2/{\alpha}\right].
\end{eqnarray*}
Here, we stress that we can take the limit $m\searrow 0$ for a fixed $\Lambda$. 
Clearly, we can recover the original form of the Cooper pair correlation. 
Therefore, in the same way as in Sec.~\ref{ProofClassical}, we obtain the desired bounds in Theorem~\ref{thm:quantum} as 
$$
\left|\left\langle c_{u,\uparrow}^\dagger c_{v,\downarrow}^\dagger 
c_{v',\downarrow}c_{u',\uparrow}\right\rangle\right|
\le {\rm Const.} \times\cases{\exp\left[-\tilde{\cal C}_4\alpha\beta^{-1} \log|u-u'| \right], & if $D=4$;\cr 
\exp\left[-\tilde{\cal C}_3\alpha\beta^{-1} |u-u'| \right], & if $D=3$, \cr}
$$
where $\tilde{\cal C}_3$ and $\tilde{\cal C}_4$ are some positive constants.

%%%%%%%%%%%%%%%%%%%%%%%%%%%%%%%%%%%%%%%%%%%%%%%%%
\appendix
 
\Section{Derivation of the bound~(\ref{SumBoundarycoshDeltaPhi})}
\label{Sec:DerboundSBcoshD}

As an example of $x\in\partial\Lambda$, we consider $x=(L,x^{(2)},\ldots,x^{(D)})$ because the rest can be 
treated in the same way.  

{From} (\ref{Deltavarphi}), one has 
\begin{eqnarray}
& &\cosh(\Delta \varphi)_x-1 \nonumber\\
&=&\cosh[(\tilde{\Delta}\varphi)_x-(d\varphi)_{x+e_1,x}]-1 \nonumber\\
&\le& \cosh|(\tilde{\Delta}\varphi)_x|\cosh|(d\varphi)_{x+e_1,x}|+\sinh|(\tilde{\Delta}\varphi)_x|
\sinh|(d\varphi)_{x+e_1,x}|-1 \nonumber \\
&\le& \left[1+\frac{\cosh qJ-1}{(qJ)^2}|(\tilde{\Delta}\varphi)_x|^2\right]\cosh|(d\varphi)_{x+e_1,x}|
+\frac{\sinh qJ}{qJ}|(\tilde{\Delta}\varphi)_x|\sinh|(d\varphi)_{x+e_1,x}|-1 \nonumber\\
&=&\left[\cosh|(d\varphi)_{x+e_1,x}|-1\right]+\frac{\cosh qJ-1}{(qJ)^2}|(\tilde{\Delta}\varphi)_x|^2
\cosh|(d\varphi)_{x+e_1,x}| 
\nonumber\\
&+&\frac{\sinh qJ}{qJ}|(\tilde{\Delta}\varphi)_x|\sinh|(d\varphi)_{x+e_1,x}|.
\label{decomIneqcoshDeltavarphi}
\end{eqnarray}
Consider first the contribution of the first term in the right-hand side. Note that 
\begin{eqnarray*}
\sum_{x\in\partial\Lambda:x^{(1)}=L}\left[\cosh|(d\varphi)_{x+e_1,x}|-1\right]&\le& 
\sum_{x\in\partial\Lambda:x^{(1)}=L}\sum_{m=1}^\infty\frac{1}{(2m)!}|(d\varphi)_{x+e_1,x}|^{2m}\\ 
&\le& \sum_{m=1}^\infty\frac{1}{(2m)!}\left[\sum_{x\in\partial\Lambda:x^{(1)}=L}|(d\varphi)_{x+e_1,x}|^2\right]^m.  
\end{eqnarray*}
{From} the expression (\ref{varphix}) of $\varphi_x$, 
$(d\varphi)_{x+e_1,x}=\varphi_{x+e_1,x}-\varphi_x$ with $x^{(1)}=L$ is computed as 
\begin{eqnarray*}
\varphi_{x+e_1,x}-\varphi_x&=&-\frac{2iq}{|\Lambda|}\sum_{k\in\mathcal{M}_\epsilon^K}
\frac{(e^{ik^{(1)}(L+1)}-e^{ik^{(1)}L})e^{i\hat{k}\cdot\hat{x}}e^{-ik\cdot(u+u')/2}}
{\Bigl[\sum_{i=1}^D(1-\cos k^{(i)})\Bigr]^2}\sin[k\cdot(u-u')/2]
\\
&=&\frac{4q}{|\Lambda|}\sum_{k\in\mathcal{M}_\epsilon^K}
\frac{e^{ik^{(1)}(L+1/2)} e^{i\hat{k}\cdot\hat{x}} e^{-ik\cdot(u+u')/2} }
{\Bigl[\sum_{i=1}^D(1-\cos k^{(i)})\Bigr]^2}\sin(k^{(1)}/2)\sin[k\cdot(u-u')/2]\\
&=&\frac{4q}{|\Lambda|}\sum_{k\in\mathcal{M}_\epsilon^K}\frac{(-1)^{n^{(1)}}e^{i\hat{k}\cdot\hat{x}}e^{-ik\cdot(u+u')/2}}
{\Bigl[\sum_{i=1}^D(1-\cos k^{(i)})\Bigr]^2}\sin(k^{(1)}/2)\sin[k\cdot(u-u')/2]
\end{eqnarray*}
where $\hat{k}=(k^{(2)},\ldots,k^{(D)})$, $\hat{x}=(x^{(2)},\ldots,x^{(D)})$, and 
we have used $k^{(1)}=2\pi n^{(1)}/(2L+1)$ with the integer $n^{(1)}$ for showing $k^{(1)}(L+1/2)=\pi n^{(1)}$. 
Therefore, we have 
\begin{eqnarray*}
& &\sum_{x\in\partial\Lambda \atop :x^{(1)}=L}|(d\varphi)_{x+e_1,x}|^2\\
&=&\frac{16q^2}{|\Lambda|^2}\sum_{x\in\partial\Lambda\atop :x^{(1)}=L}\sum_{p\in\mathcal{M}_\epsilon^K}
\frac{(-1)^{m^{(1)}}e^{-i\hat{p}\cdot\hat{x}}e^{ip\cdot(u+u')/2}}
{\Bigl[\sum_{i=1}^D(1-\cos p^{(i)})\Bigr]^2}\sin(p^{(1)}/2)\sin[ p\cdot(u-u')/2]\\
&\times&\sum_{k\in\mathcal{M}_\epsilon^K}\frac{(-1)^{n^{(1)}}e^{i\hat{k}\cdot\hat{x}}e^{-ik\cdot(u+u')/2}}
{\Bigl[\sum_{i=1}^D(1-\cos k^{(i)})\Bigr]^2}\sin(k^{(1)}/2)\sin[k\cdot(u-u')/2]\\
&=&\frac{16q^2}{(2L+1)|\Lambda|}\sum_{p^{(1)}\atop :\epsilon\le|p^{(1)}|\le K}\sum_{k\in\mathcal{M}_\epsilon^K}
\frac{(-1)^{m^{(1)}}e^{ip^{(1)}(u^{(1)}+{u'}^{(1)})/2}}
{[(1-\cos p^{(1)})+\sum_{i=2}^D(1-\cos k^{(i)})]^2}\\
&\times&\sin(p^{(1)}/2)\sin[p^{(1)}(u^{(1)}-{u'}^{(1)})/2+\hat{k}\cdot(\hat{u}-\hat{u}')/2]\\
&\times&\frac{(-1)^{n^{(1)}}e^{-ik^{(1)}(u^{(1)}+{u'}^{(1)})/2}}
{\Bigl[\sum_{i=1}^D(1-\cos k^{(i)})\Bigr]^2}\sin(k^{(1)}/2)\sin[k\cdot(u-u')/2],
\end{eqnarray*}
where $p^{(1)}=2\pi m^{(1)}/(2L+1)$ with the integer $m^{(1)}$. 
Here, the integers, $m^{(1)}$ and $n^{(1)}$, take both of 
even and odd values, and the even and odd integers yield a contribution with the same absolute value 
but opposite sign in the limit $L\rightarrow\infty$. Consequently, we obtain 
$$
\lim_{L\rightarrow\infty}\sum_{x\in\partial\Lambda}|(d\varphi)_{x+e_{\bot},x}|^2=0, 
$$
where $e_{\bot}$ is the unit normal vector to the boundary $\partial\Lambda$ 
at the site $x\in\partial\Lambda$. 
Therefore, the corresponding contribution is vanishing in the limit. 

Next, consider the second term in the right-hand side of the last equality in (\ref{decomIneqcoshDeltavarphi}). 
Note that
\begin{eqnarray*}
\sum_{x\in\partial\Lambda}|(\tilde{\Delta}\varphi)_x|^2[\cosh|(d\varphi)_{x+e_\bot,x}|-1]
&\le&\sum_{x\in\partial\Lambda}|(\tilde{\Delta}\varphi)_x|^2
\sum_{y\in\partial\Lambda}[\cosh|(d\varphi)_{y+e_\bot,y}|-1]\\
&=&\sum_{x\in\partial\Lambda}|(\tilde{\Delta}\varphi)_x|^2
\sum_{y\in\partial\Lambda}\sum_{m=1}^\infty \frac{1}{(2m)!}|(d\varphi)_{y+e_\bot,y}|^{2m}\\
&\le&\sum_{x\in\partial\Lambda}|(\tilde{\Delta}\varphi)_x|^2
\sum_{m=1}^\infty \frac{1}{(2m)!}\left[\sum_{y\in\partial\Lambda}|(d\varphi)_{y+e_\bot,y}|^2\right]^m.  
\end{eqnarray*}
{From} the above observation, it is sufficient to evaluate  
$$
\sum_{x\in\partial\Lambda}|(\tilde{\Delta}\varphi)_x|^2. 
$$
{From} the expression (\ref{tildeDeltavarphi}), we have 
$$
\sum_{x\in\partial\Lambda}|(\tilde{\Delta}\varphi)_x|^2\le 
\sum_{x\in\Lambda}|(\tilde{\Delta}\varphi)_x|^2=\frac{16q^2}{|\Lambda|}\sum_{k\in\mathcal{M}_\epsilon^K}
\frac{\sin^2[k\cdot (u-u')/2]}{\Bigl[\sum_{i=1}^D(1-\cos k^{(i)})\Bigr]^2}.
$$
Because of $D\ge 3$, this right-hand side is bounded uniformly in $\Lambda$ for fixed $u, u'$ and $\epsilon$. 

The third term in (\ref{decomIneqcoshDeltavarphi}) can be evaluated as 
\begin{eqnarray*}
\sum_{x\in\partial\Lambda}|(\tilde{\Delta}\varphi)_x|\sinh|(d\varphi)_{x+e_\bot,x}|&\le&
\sqrt{\sum_{x\in\partial\Lambda}|(\tilde{\Delta}\varphi)_x|^2}\sqrt{\sum_{y\in\partial\Lambda}\sinh^2|(d\varphi)_{y+e_\bot,y}|}\\
&\le&\sqrt{\sum_{x\in\Lambda}|(\tilde{\Delta}\varphi)_x|^2}
\sqrt{\sum_{y\in\partial\Lambda}\sinh^2|(d\varphi)_{y+e_\bot,y}|}
\end{eqnarray*} 
by using the Schwarz inequality. Further, we have 
\begin{eqnarray*}
\sum_{y\in\partial\Lambda}\sinh^2|(d\varphi)_{y+e_\bot,y}|
&=&\sum_{y\in\partial\Lambda}\sum_{m=0}^\infty \sum_{n=0}^\infty 
\frac{1}{(2m+1)!}\frac{1}{(2n+1)!}|(d\varphi)_{y+e_\bot,y}|^{2(m+n)+2}\\
&\le&\sum_{m=0}^\infty \sum_{n=0}^\infty 
\frac{1}{(2m+1)!}\frac{1}{(2n+1)!}\left[\sum_{y\in\partial\Lambda}|(d\varphi)_{y+e_\bot,y}|^2\right]^{(m+n)+1}.
\end{eqnarray*} 
As we showed as in the above, the right-hand side is vanishing in the limit $L\rightarrow\infty$. 
Thus, the corresponding contribution is vanishing in the limit. 

Putting these together, we obtain the desired bound (\ref{SumBoundarycoshDeltaPhi}).

%%%%%%%%%%%%%%%%%%%%%%%%%%%%%%%%%%%%%%%%%%%%%%%%%%%%%%%%%%
\Section{Proof of Lemma~\ref{lem:TrPUUP}} 
\label{Sec:ProofLemTrPUUP}

To begin with, we write 
$$
\mathcal{Z}:={\rm Tr}\; P^{2\ell}. 
$$
It is sufficient to show 
\begin{equation}
\label{PUboundP}
\frac{1}{\mathcal{Z}}{\rm Tr}\; (PU)^\ell (U^\ast P)^\ell\le 1.
\end{equation}

Consider the set of quantities, 
$$
\frac{1}{\mathcal{Z}}\left|{\rm Tr}\; PU_1PU_2\cdots PU_{\ell-1}P\cdot PU_{\ell+1}PU_{\ell+2}P\cdots PU_{2\ell-1}P\right|,
$$
where $U_i$, $i=1,2,\ldots,\ell-1,\ell+1,\ldots,2\ell-1$, are an unitary operator, and take a value 
$U_i\in\{U,U^\ast,1\}$. Clearly, the set contains the quantity of the left-hand side of (\ref{PUboundP}).  
We write 
$$
a_{\rm max}:=\frac{1}{\mathcal{Z}}\max_{U_i\in\{U,U^\ast,1\}}\{\left|{\rm Tr}\; PU_1PU_2\cdots 
PU_{\ell-1}P\cdot PU_{\ell+1}PU_{\ell+2}P\cdots PU_{2\ell-1}P\right|\}
$$
for the maximum value. 

We set  
$$ 
\frac{1}{\mathcal{Z}}\left|{\rm Tr}\; PU_1PU_2\cdots 
PU_{\ell-1}P\cdot PU_{\ell+1}PU_{\ell+2}P\cdots PU_{2\ell-1}P\right|=a_{\rm max}. 
$$
Using the property of the trace and the Schwarz inequality, one has 
\begin{eqnarray*}
a_{\rm max}&=&\frac{1}{\mathcal{Z}}\left|{\rm Tr}\; PU_1PU_2\cdots 
PU_{\ell-1}P\cdot PU_{\ell+1}PU_{\ell+2}P\cdots PU_{2\ell-1}P\right|\\
&=&
\frac{1}{\mathcal{Z}}\left|{\rm Tr}\; P^2U_1PU_2\cdots 
PU_{\ell-1}\cdot P^2U_{\ell+1}PU_{\ell+2}P\cdots PU_{2\ell-1}\right|\\
&\le&\sqrt{\frac{1}{\mathcal{Z}}\left|{\rm Tr}\; P^2U_1PU_2\cdots P U_{\ell-2}P \cdot 
P U_{\ell-2}^\ast P \cdots P U_2^\ast P U_1^\ast P^2\right|}\; a_1^{1/2},
\end{eqnarray*}
where we have written 
$$
a_1=\frac{1}{\mathcal{Z}}{\rm Tr}\; 
P U_{2\ell-2}^\ast P \cdots P U_{\ell+2}^\ast P U_{\ell+1}^\ast P^2 \cdot P^2U_{\ell+1}PU_{\ell+2}P\cdots U_{2\ell-2}P.
$$
Similarly, 
\begin{eqnarray*}
& &
\frac{1}{\mathcal{Z}}\left|{\rm Tr}\; P^2U_1PU_2\cdots P U_{\ell-2}P \cdot 
P U_{\ell-2}^\ast P \cdots P U_2^\ast P U_1^\ast P^2\right|\\
&=&\frac{1}{\mathcal{Z}}\left|{\rm Tr}\; P^3U_1PU_2\cdots P U_{\ell-2}\cdot   
P^2 U_{\ell-2}^\ast P \cdots P U_2^\ast P U_1^\ast P\right|\\
&\le& \sqrt{\frac{1}{\mathcal{Z}}\left|{\rm Tr}\; P^3U_1PU_2\cdots U_{\ell-3}P \cdot
P U_{\ell-3}^\ast \cdots U_2^\ast P U_1^\ast P^3 \right|}\times a_2^{1/2}.
\end{eqnarray*} 
Therefore, we have 
$$
a_{\rm max}\le \left[\frac{1}{\mathcal{Z}}\left|{\rm Tr}\; P^3U_1PU_2\cdots U_{\ell-3}P \cdot
P U_{\ell-3}^\ast \cdots U_2^\ast P U_1^\ast P^3 \right|\right]^{1/4}a_2^{1/4}a_1^{1/2}.
$$
By repeating this procedure, we obtain 
$$
a_{\rm max}\le \left(\frac{1}{\mathcal{Z}}{\rm Tr}\; P^\ell \cdot P^\ell \right)^{1/{2^{\ell-1}}}
a_{\ell-1}^{1/{2^{\ell-1}}}\cdots a_2^{1/{2^2}} a_1^{1/2}
=a_{\ell-1}^{1/{2^{\ell-1}}}\cdots a_2^{1/{2^2}} a_1^{1/2}\le (a_{\rm max})^{1-2^{-\ell}}.
$$
This implies $a_{\rm max}\le 1$. 
 
%%%%%%%%%%%%%%%%%%%%%%%%%%%%%%%%%%%%%%%%%%%%%%%%%%%%%%%%%%%%%%%%%%%%%%%%%%%
\bigskip\bigskip\bigskip

\noindent
{\bf Acknowledgements:} We would like to thank Hal Tasaki for helpful discussions. 
%Peter Fulde, Hosho Katsura, Masaaki Shimozawa, 
% and Masafumi Udagawa. 
YT was partly supported by JSPS/MEXT Grant-in-Aid for Scientific
Research (Grant No. 26800177) and by Grant-in-Aid for
Program for Advancing Strategic International Networks to
Accelerate the Circulation of Talented Researchers (Grant No. R2604) ``TopoNet." 
%%%%%%%%%%%%%%%%%%%%%%%%%%%%%%%%%%%%%%%%%%%%%%%%%%%%%%%%%%%%%
%\newpage

\end{document}